\documentclass[aps,prd,10pt,twocolumn,nofootinbib,floatfix]{revtex4}
 
\usepackage{epsfig}
\usepackage{bm}
\usepackage{latexsym}
\usepackage{natbib}
\usepackage{url}
\usepackage{dcolumn}
\usepackage{color}
\usepackage{amsfonts,amssymb,amsmath}
\usepackage{graphicx,epsfig}
\usepackage{psfrag}
\usepackage{subfigure}
\usepackage{hyperref}
\hypersetup{colorlinks=true}
\usepackage{mathtools}
\usepackage{enumitem}
\usepackage{float}
\usepackage{mathrsfs}
\usepackage[dvipsnames]{xcolor}
\usepackage[compat=1.1.0]{tikz-feynman}
\hypersetup{  colorlinks,  citecolor=red,  linkcolor=red,  urlcolor=blue}

\usepackage{comment}

\begin{document}

\title{GRB221009A gamma-ray events from non-standard neutrino self-interactions}

\author{
Mansi Dhuria$^{a,}$\footnote{Mansi.dhuria@sot.pdpu.ac.in}
}

\affiliation{
$^a$ Department of Physics, School of Energy Technology, Pandit Deendayal Energy University (PDEU), Gandhinagar-382426, Gujarat, India
}

\begin{abstract}
\noindent 

The flux of high-energy astrophysical neutrinos observed by the present generation of neutrino detectors has already indicated a few hints of new physics beyond the Standard Model. In this work, we show that high-energy gamma-ray observations can also be considered as a complementary probe for unveiling the source of high-energy astrophysical neutrino events and new physics. Recently, the LHAASO collaboration has reported $
{\cal O}$(5000) gamma-ray events in the energy range between 0.5 TeV -18 TeV from gamma-ray burst GRB 221009A within 2000 seconds after the initial outburst.  We showed that attenuated high-energy gamma rays can be produced from the interaction of astrophysical neutrinos with CMB neutrinos through non-standard self-interaction of neutrinos mediated by light scalar bosons. The non-standard interaction of neutrinos recently took a lot of attention in cosmology for its role in reducing Hubble tension. We have constrained the parameter space of non-standard self-interacting neutrinos from the flux of photons observed by LHAASO and showed consistency of the same with the resulting parameter space from Hubble tension requirements and other recent constraints from laboratory/cosmology.
 
\end{abstract}

\maketitle

\section{Introduction}
The results from various cosmological and astrophysical observations have convinced us that neutrinos play a significant role in multi-messenger astronomy as one of the messengers used to study exotic astrophysical phenomena as well as in understanding the evolution of the universe. In the standard model (SM) of particle physics, neutrinos are considered to interact very weakly through the weak force, thus making them challenging to detect. However, recently, there has been a lot of debate about the possibility of moderately strong non-standard interaction of neutrinos with each other through a new mediator,  typically known as self-interaction of neutrinos ($\nu$SI). It plays an important role in reducing the Hubble tension~\cite{Lancaster:2017ksf,Oldengott:2017fhy,forastieri2019cosmological, Das:2020xke,brinckmann2021self,choudhury2021updated}, allowing KeV sterile neutrino as viable Dark matter (DM) candidate~\cite{DeGouvea:2019wpf,Kelly:2020pcy,Benso_2022,Dhuria:2023yrw} and supernova
neutrino emission~\cite{Chen:2022kal}. As the new mediator required to include self-interaction between neutrinos naturally invokes the existence of physics beyond the Standard Model (BSM), one can study its implications in explaining other issues in astrophysics and cosmology that might include BSM physics. In this work, we have considered the presence of self-interaction of neutrinos in explaining the flux of high-energy photons obtained from gamma-ray bursts.

Recently, on 09 October 2022, a prodigious bright gamma-ray burst (dubbed GRB221009A) was first recorded by the
Burst Alert Telescope (BAT) on the Swift satellite~\cite{2022GCN.32635....1K} and later confirmed by Fermi Gamma-ray burst Monitor (GBM)~\cite{2022GCN.32636....1V,2022GCN.32642....1L} and Fermi-LAT~\cite{2022GCN.32637....1B,2022GCN.32658....1P} at redshift  $z_0 \sim 0.15$~\cite{2022GCN.32648....1D, 2022GCN.32686....1C}. The extremely energetic
gamma rays emitted from the cosmic burst were detected by Large High Altitude Air Shower Observatory (LHAASO)~\cite{2022GCN.32677....1H} and Carpet-2 experiment~\cite{2022ATel15669....1D}. More specifically, the Square kilometer array (KM2A) of LHAASO (\cite{2022GCN.32677....1H}) reported the observation of around 5000 very-high-energy photons with energy up to 18 TeV in a 2000 sec time window while the carpet-2 experiment reported the claim for detection of 251 TeV photon-like shower events~\cite{2022ATel15669....1D}.  The reported redshift corresponds to a co-moving distance of around 643 Mpc from the Earth. The observation of these events is quite astonishing as the flux of such photons would be severely attenuated because of the pair production of electrons and positrons via interaction with extra-galactic background light (EBL)($\gamma + \gamma_{\rm EBL}\rightarrow e^{-} e^{+}$).  Therefore, the photons would hardly arrive on the earth. This has speculated many proposals by invoking BSM Physics. The observation of such events has been explained from sterile neutrino decay~\cite{Cheung:2022luv,Smirnov:2022suv,Brdar:2022rhc,Guo:2023bpo}, scalar decay \cite{Balaji:2023nbn}, Lorentz invariance violation~\cite{Finke:2022swf,Vardanyan:2022ujc}, axion-photon conversion \cite{Baktash:2022gnf,Galanti:2022pbg,Galanti:2022xok,Carenza:2022kjt,Lin:2022ocj,Troitsky:2022xso,Nakagawa:2022wwm,Huang:2022udc,Wang:2023okw,Bernal:2023rdz,Troitsky:2023uwu,Li:2023rhj} and inverse Compton mechanism~\cite{Zhang:2022lff,AlvesBatista:2022kpg,Gonzalez:2022opy} etc. 

It has been observed that in addition to photons, neutrinos can also be produced through the decay of kaons and muons emitted during gamma-ray bursts. The IceCube collaboration has also performed
dedicated searches for co-relating some of the GRB events with diffuse extra-galactic neutrino background of very high energy neutrinos~\cite{IceCube:2014jkq,IceCube:2022rlk} and also constrained the time-integrated flux of neutrinos from GRB221009A at Earth~\cite{Ai:2022kvd, Murase:2022vqf}. It has been argued in~\cite{Cheung:2022luv,Smirnov:2022suv,Brdar:2022rhc,Guo:2023bpo} that neutrinos emitted from GRB can convert into sterile neutrinos through mixing or dipole interaction. The produced sterile neutrino can then decay into photons through the same process after propagating for a long distance so that EBL would not significantly attenuate this secondary photon flux. Thus, the observation of photons can be explained through the decay of sterile neutrino. However, in this case, the flux of photons would depend on the mixing angle between sterile and active neutrinos. The mixing angle required to explain a non-zero number of photons ($N_{\gamma}>1$) would be very high (0.1-1), which is ruled out by experiments.  Further, the resulting parameter space also gets ruled out by various astrophysical and cosmological observations~\cite{Guo:2023bpo}. 
                                                                    
In this work, we consider the possibility of producing such events from a direct scattering of active neutrinos with cosmic microwave background (CMB) neutrinos rather than the decay of sterile neutrinos produced through mixing with active neutrinos. The high-energy neutrinos produced during GRB can scatter with CMB neutrinos through newly proposed self-interactions of neutrinos mediated by scalar bosons. If the mediator also interacts with a corresponding leptonic partner of the neutrino in a specific model of BSM physics, the interaction of high-energy neutrinos with CMB neutrinos can produce high-energy photons and background CMB photons at a one-loop level. Even though the loop-level cross-section will be suppressed than the cross-section for elastic scattering of high energy neutrinos with CMB neutrinos by a small order ($10^{-2}-10^{-3})$, there will be a non-zero probability of producing photons from such scatterings. We have observed that the mean free path for the scattering can be such that the astrophysical high energy photons would be produced near the surface of the earth and hence they will not be attenuated by EBL. More interestingly, some of the parameter space required to obtain a relevant number of gamma-ray events is also compatible with the parameter space required to resolve/reduce Hubble tension. In fact, results are also compatible with the allowed region of $(g-2)_{\mu}$ constraints and safe from other astrophysical and cosmological constraints.

The plan of the rest of the paper is as follows: In \$II, we motivate the role of self-interacting neutrinos in cosmology, specifically for alleviating Hubble tension. In \$II(a) and \$II(b), we discuss the toy model of particle physics by involving the interaction of the scalar mediator with neutrino and its leptonic partner. Then we explain the possibility of producing astrophysical photons from the scattering of high-energy astrophysical neutrinos with CMB neutrinos. In \$III, we estimate the effect of such interactions on the astrophysical flux of photons generated from loop-level scattering. We also discuss that some of the resulting parameter space is also compatible with constraints from Hubble tension requirement and $(g-2)_\mu$ discrepancy. Finally, in \$IV, we discuss our results with interesting conclusions and future directions.

\section{Self-interacting neutrinos}
The self-interacting neutrinos refer to a scenario where neutrinos interact strongly with each other secretly through a new interaction mediated by either scalar or vector boson. This will allow neutrinos to remain in thermal equilibrium with each other till later times. Thus, the epoch of neutrino decoupling gets delayed until even close to the onset of matter-radiation equality.  The effect of the same on the CMB power spectrum has been analyzed in detail in literature~\cite{Lancaster:2017ksf,Oldengott:2017fhy,Das:2020xke,forastieri2019cosmological,brinckmann2021self,choudhury2021updated} in the context of Hubble tension.  Below, we have briefly discussed the impact of strong self-interactions in increasing the value of the Hubble constant measured by CMB in the context of the $\Lambda$CDM model. 

The position of CMB multiple for a particular mode $k$ can is given by~\cite{Das:2020xke}
\begin{equation}
\label{eq:l}
    l\approx \frac{(m\pi-\phi_\nu)}{\theta_\ast},~~{\rm with}~{\theta_\ast}=\frac{r^\ast_s}{D^\ast_A} \,,
\end{equation}
where $m\pi$ corresponds to the position of peaks, $\phi_\nu$  corresponds to the phase shift, $D^\ast_A$ is the distance between the surface of the last scattering and today, and $r^\ast_s$ corresponds to the radius of the sound horizon at the epoch of recombination. The quantities $D^\ast_A$ and $r^\ast_s$ can be expressed as a function of the Hubble parameter $H(z)$ as~\cite{Das:2020xke}:
$D^\ast_A = \int_{0}^{z^\ast}\frac{1}{H(z)} \,dz $
     and $ r^\ast_s = \int_{z^\ast}^{\infty}\frac{c_s(z)}{H(z)} \, dz,$ where $c_s(z) \approx 1/\sqrt{3}$ is the speed of sound in the baryon-photon plasma. The phase shift ($\phi_\nu$) depends on the ratio of free-streaming neutrino energy density to
the total radiation energy density~\cite{Das:2020xke}.
The decrease in the number density of free-streaming neutrino will decrease the phase shift $\phi_\nu$, which further leads to a shift in the position of the CMB multiple towards a high $l$ value. This change can be avoided by increasing the value of $\theta_\ast$, which can be achieved either by decreasing the value of $D^\ast_A$ while keeping $r^\ast_s$  unchanged or increasing the value of $r^\ast_s$ while keeping $D^\ast_A$  unchanged. In the standard cosmological model, the evolution of Hubble constant evolves with redshift $z$ is given by $H(z) = H_0 \sqrt{\Omega_r (1+z)^4 + \Omega_m (1+z)^3 + \Omega_\Lambda}$, with $\Omega_m$, $\Omega_r$ and $\Omega_\Lambda$  being the fraction of the energy density acquired by radiation, matter, and vacuum respectively. If we slightly increase the value of $H_0$ and  $\Omega_\Lambda$ such that there is an increase in the value of $H(z)$ at low redshift while there is negligible change for $H(z)$ at high redshifts, one can decrease $D^\ast_A$ in such a way that the position of observed CMB multipoles will not be changed. 
In this way, the presence of self-interacting neutrinos impels a higher value of $H_0$ and reduces the discrepancy between the value of $H_0$ measured by CMB and low-redshift observations.

\subsection{The model}
 The minimal model of self-interaction typically involves neutrinophilic interaction given by ${\cal L} \supset  {g_\nu}  \nu_i \nu_i$, where $i= e,\mu,\tau$ corresponds to different flavors of active neutrinos. As IceCube has searched
only for track-like events from GRB221009A, we consider self-interaction only between muon flavors of neutrinos. Interestingly,  the interaction of scalar with muons is also considered in the literature for resolving the discrepancy between the theoretical and experimental values of $(g-2)_{\mu}$~\cite{Liu:2021kug}. Therefore, in this work, we consider a model in which the real singlet scalar at low energies couples both to muon  
neutrinos as well as muons. The interaction couplings are given as:
\begin{equation}
{\cal L } \supset g_{\mu }\phi {\bar \mu}{\mu} + g_{\nu_\mu}\phi {\nu_\mu}{\nu_\mu}.
\end{equation}
Assuming the Majorana nature of neutrinos, we have used Weyl notation to represent neutrino coupling to the scalar. For muons, we are considering Dirac notation to represent its coupling with the scalar boson. The interaction coupling $g_{\nu_\mu}$ can be generally estimated from the mechanism of neutrino mass generation, while $g_{\mu }$ can be obtained from the non-renormalizable coupling involving Higgs field and a new scalar field. As we are not interested in the UV completion of the model, we just assume that both the couplings can be the same or different depending on the particular model used to generate such couplings.

\subsection{Scattering of self-interacting neutrinos with CMB neutrinos}
\begin{figure}[htbp]
\centering
  \includegraphics[width=0.35\textwidth, height =0.2\textwidth]{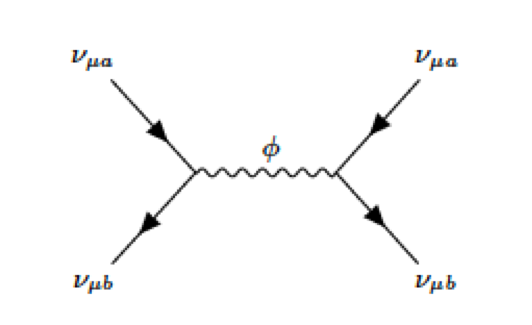}
     \caption{Feynman diagram for scattering of high energy astrophysical neutrinos with CMB neutrinos. Here, ${\nu_{\mu a} }$ and ${ \nu_{\mu b} }$ correspond to high-energy astrophysical neutrinos and background CMB neutrinos respectively.}
     \label{fig:elasticscat}
\end{figure}

 Neutrinos are often considered to be astounding ``messengers" because of their capability to carry insight into various astrophysical processes and events happening in the universe. In SM, neutrinos can traverse large distances on their way to Earth because the probability of an interaction between neutrinos and matter is very small. In the presence of new non-standard interaction of neutrinos, the high energy neutrinos emitted from various astrophysical objects such as gamma-ray bursts, galaxy sources, etc. might scatter with other new particles while traveling to Earth. Thus, there is a possibility that the mean free path of neutrino will become smaller. In the case of self-interacting neutrinos, the mean free path of neutrinos can be affected due to the secret self-interaction of astrophysical neutrinos with the cosmic neutrino background. The high-energy astrophysical neutrinos can scatter strongly with CMB neutrinos while traveling to Earth and impact the flux of astrophysical neutrinos observed from various sources. In other words, the flux of high-energy astrophysical neutrinos obtained from various sources can provide complementary probes to constrain the strength of non-standard interaction of neutrinos. The effect has been analyzed in \cite{Hooper:2023fqn,Doring:2023vmk} by analyzing constraints on the self-interacting neutrino coupling from the flux of astrophysical neutrino point sources obtained from NGC and IceCube results. The Feynman diagram representing the s-channel scattering of astrophysical neutrinos (${\nu_{\mu a}}$) with CMB neutrinos (${\nu_{\mu b} }$) has been shown in fig~\ref{fig:elasticscat}. 

\begin{figure}[htbp]
\centering
  \includegraphics[width=0.35\textwidth, height =0.27\textwidth]{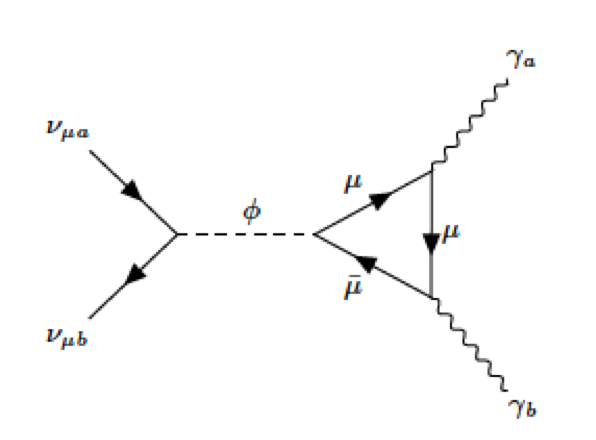}
     \caption{Feynman diagram for scattering of high energy astrophysical neutrinos (${\nu_{\mu a} }$) with CMB neutrinos (${\nu_{\mu b} }$) into high energy astrophysical photons $(\gamma_a)$ and CMB background photons $(\gamma_b)$.}
     \label{fig:inelasticscat}
\end{figure}

In this work, we have proposed that the s-channel scattering of astrophysical neutrinos can also produce high-energy gamma rays at the radiative level if the mediator interacts with both muon neutrinos and muons. Thus, the flux of high-energy gamma rays obtained from various sources can also be considered to constrain the strength of non-standard interactions of neutrinos. The one-loop level Feynman diagram for producing photons from radiative s-channel scattering of astrophysical neutrinos (${\nu_{\mu a}}$) with CMB neutrinos (${\nu_{\mu b} }$) has been shown in fig~\ref{fig:inelasticscat}. 

We will analyze constraints on the strength of non-standard interaction of neutrinos from the flux of high energy gamma-ray events observed by GRB221009A. We will also analyze if the self-interaction coupling required to produce such high-energy photons is consistent with the values required to address the Hubble tension and other recent laboratory/cosmological constraints.

\section{Impact on astrophysical gamma-ray flux from GRB221009A}
As discussed earlier, the LHAASO collaboration has recently reported an extremely bright and long-duration Gamma Ray Burst, named GRB221009A~\cite{2022GCN.32635....1K}. They have detected ${\cal O}(5000)$ events of photons with energies ranging from 0.5 TeV to 18 TeV within a time window of 2000 at redshift  $z=0.15$. The violent reactions around GRB normally produce a large number of pions or kaons which can further decay into photons and neutrinos. As the photon will interact with background photons to annihilate into electron-positron pairs, the flux of astrophysical neutrinos will be significantly attenuated. Thus, the detection can not be explained due to extragalactic background light coming from GRB events. Interestingly, the observed flux of high and very-high-energy photons can be obtained from neutrinos emitted during GRB events through the interaction of emitted high-energy astrophysical neutrinos with CMB neutrinos. Thus, GRB events would provide a unique opportunity to probe the non-standard interaction of neutrinos. 

In this section, we calculate the probability of producing high-energy gamma rays from the scattering of high-energy astrophysical neutrinos with CMB neutrinos. The unattenuated $\gamma$ flux of GRB 221009A obtained by extrapolating the flux measured by FermiLAT  in the energy range (0.1 - 1)~GeV to higher energies (around TeV) is given by~\cite{Smirnov:2022suv}
\begin{equation}
\phi^0_{\gamma}(E_\gamma) = \frac{2.1\times 10^{-6}}{{\rm cm}^2 {\rm s}^{-1} {\rm TeV}} \left(\frac{E_\gamma}{\rm TeV}\right)^{-1.87 \pm 0.04}.
\end{equation}
The emission of neutrinos from GRB221009A has been analyzed by complementary experiments such as Icecube. 
The non-observation of track-like neutrino events in the energy range 0.8 TeV - 1 PeV has set constraints on neutrino fluence $E^2_\nu \phi^{\rm int}_{\nu}  \le 3.9 \times 10^{-5} ~ {\rm TeV}~{\rm cm}^{-2}$ ~\cite{Ai:2022kvd, Murase:2022vqf}. As neutrinos would mainly be emitted from muon and kaon decay, they would mostly consist of astrophysical muon neutrinos ($\nu_{\mu a}$). Therefore, the  ratio of the flux of neutrinos to the flux of unattenuated gamma rays will be given by
\begin{equation}
\label{rnugamma}
r_{\nu \gamma} = \frac{\phi_{\nu_{\mu a}}}{\phi^0_\gamma(E_\gamma)}.
 \end{equation}
By dividing the neutrino fluence with a long period ($\Delta \tau \sim 600$ sec) of intense gamma-ray emission, one gets the ratio of the fluxes  $r_{\nu \gamma} \lesssim 3 \times 10^{-2}$~\cite{Smirnov:2022suv}.

As shown in fig.~\ref{fig:elasticscat}, the high energy neutrinos emitted from muon/kaon decay produced during GRB at redshift z=0.15 can scatter with CMB neutrinos. The optical depth of neutrino would be given by:
\begin{equation}
    \tau_{\nu_\mu} = \frac{\lambda_{\nu_\mu}}{d},
    \end{equation}
 where ${\lambda_{\nu_\mu}}$ corresponds to total mean free path of neutrinos. The mean free path of neutrinos annihilating into gamma rays will be given by  ${\lambda_{{\nu_\mu}\rightarrow \gamma}} = {{\cal BR}({\nu_{\mu a} \nu_{\mu b} \rightarrow \gamma_a \gamma_b})} {\lambda_{\nu_\mu}}$. 
 
 Thus, the probability of receiving gamma rays on earth from the scattering of astrophysical neutrinos with CMB neutrinos in the distance interval [x, x+dx] will be given by:
 \begin{equation}
 \label{eq:probability}
 e^{-x/{{\lambda_{\nu_\mu \rightarrow \gamma}} }}~\frac{dx}{{\lambda_{\nu_\mu\rightarrow \gamma}} }~e^{-(d-x)/\lambda_{\gamma}},
\end{equation}
where  ${\lambda_{\nu_\mu\rightarrow \gamma}}$ is the mean free path for the scattering of neutrinos into gamma rays and $\lambda_{\gamma}$ corresponds to the mean free path of gamma rays. Multiplying eq.~(\ref{eq:probability}) by neutrino flux 
and integrating over $x$, the secondary gamma-ray flux from the neutrino scattering will be given by:
 \begin{equation}
 \phi^{\gamma}_{\nu_\mu} = \phi_{\nu_\mu} \frac{1}{\left({\lambda_{\mu \rightarrow \gamma}}/{\lambda_\gamma}\right)-1} \left[e^{-d/\lambda_{\mu}} -e^{-d/\lambda_{\gamma}} \right]
\end{equation}
For $\phi_{\nu_{\mu a}} = r_{\nu \gamma} \times \phi^0_\gamma(E_\gamma) = 0.03~\phi^0_\gamma(E_\gamma) $ using eq.~(\ref{rnugamma}), the  secondary gamma ray flux will be:
 \begin{equation}
 \label{eq:fluxfin}
 \phi^{\gamma}_{\nu_\mu} =  0.03 \ \frac{\phi^0_\gamma}{\left({\lambda_{\mu \rightarrow \gamma}}/{\lambda_{\gamma}}\right)-1} \left[e^{-d/\lambda_{\mu \rightarrow \gamma}} -e^{-d/\lambda_{\gamma}} \right]
\end{equation}
In the above expression, the second exponential factor corresponds to the gamma-ray flux produced directly in GRB and can be ignored for $\lambda_{\mu \rightarrow \gamma} \sim d \approx 10^{27}$ cm. Hence, there is a possibility that the gamma-ray flux produced from the scattering of neutrinos will not be exponentially attenuated as compared to the gamma-ray flux produced directly produced from GRB. In forthcoming subsections, we numerically estimate the mean free path for astrophysical neutrinos and present results related to the flux of astrophysical gamma rays produced from the scattering of muon neutrinos.

\subsection{Mean free path of neutrinos}
\begin{figure}[htbp]
\centering
\hskip -0.2in \includegraphics[width=0.5\textwidth, height =0.3\textwidth]{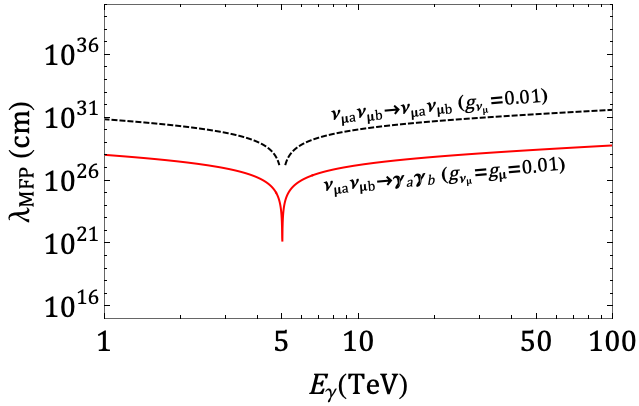}
     \caption{The black dashed line corresponds to the mean free path of neutrino for the elastic scattering of an incident astrophysical neutrino with CMB neutrinos for interaction coupling $g_{\nu_\mu}=0.01$. The red solid line shows the mean free path of neutrino for $g_{\nu_\mu}= g_{\mu}=0.01$ when the interaction of an incident astrophysical neutrino with CMB neutrinos produces gamma rays. The dip in the curve shows the resonance enhancement of the cross-section at $s \approx 4 m^2_\phi$.}
     \label{fig:mfp}
\end{figure}

The mean free path of neutrinos emitted from astrophysical sources can be calculated from the interaction rate of incident neutrinos with the background CMB neutrinos. The value of ${\lambda_{\nu_\mu\rightarrow \gamma}}$ will be given by~\cite{Doring:2023vmk}
\begin{equation}
\label{eq:mfp}
{\lambda_{\nu_\mu\rightarrow \gamma}} = \frac{1}{\Gamma({\nu_{\mu a} \nu_{\mu b} \rightarrow   \gamma_a \gamma_b})},
\end{equation}
where $\Gamma({\nu_{\mu a} \nu_{\mu b} \rightarrow   \gamma_a \gamma_b})$ is the interaction rate of the incident neutrino with CMB neutrinos. The cross-section for the production of gamma rays from the scattering of astrophysical neutrinos with CMB neutrinos (shown in Feynman diagram given in fig.~\ref{fig:inelasticscat}) is given by~\cite{Duerr:2015vna,FileviezPerez:2019rcj}
\begin{eqnarray}
\label{eq:crossec}
    && \sigma({\nu_{\mu a} \nu_{\mu b} \rightarrow   \gamma_a \gamma_b}) = \frac{81 \alpha^2 s}{4 \pi^3} \frac{(g_{\mu }g_{\nu_\mu })^2 }{(s - m^2_\phi)^2 + m^2_\phi \Gamma^2_\phi } \nonumber\\
    && \times \left|1+ \sum_f  Q^2_\mu m^2_{\mu} C^\gamma_0 \right|^2,
\end{eqnarray}
where scalar Passarino–Veltman function $C^\gamma_0$ is given by,
\begin{equation}
\label{eq:crossec1}
    C^\gamma_0 (s, m_\mu) = \frac{1}{2 s} {\rm ln^2}\left(\frac{\sqrt{1-  {4 m^2_\mu}/s}-1}{\sqrt{1- {4 m^2_\mu}/s}+1} \right).
\end{equation}
By using the expression of cross section given in eq.~(\ref{eq:crossec}), the thermal interaction rate is given by~\cite{Doring:2023vmk}
\begin{equation}
\label{eq:intrate}
{\hskip -0.07in} \Gamma ({\nu_{\mu a} \nu_{\mu b} \rightarrow   \gamma_a \gamma_b}) = \int \frac{d^3p}{(2 \pi)^3} f_i (\vec {p_i}) v_{\rm M o l} \sigma(s(E_\nu, {\vec p})).
 \end{equation}
Today, the CMB neutrino background has a thermal distribution with total number density $n_{\rm tot} \approx 340 {\rm cm}^{-3}$ and temperature $T_\nu = 1.9$ K. Given this, the background CMB neutrino can be considered as non-relativistic with $m_{\nu_{\mu}} > T_\nu$. In this case, the  center of mass energy $s$  becomes independent of the momentum and we can get $s= \sqrt{2 m_{\nu_\mu} E_{\nu_{\mu a}}}$ and  $v_{\rm M o l}=1$. Using this, the integral can be easily solved in a lab frame and the interaction rate for scattering with non-relativistic background neutrinos reduces to
\begin{equation}
\label{eq:intrate1}
\Gamma ({\nu_{\mu a} \nu_{\mu b} \rightarrow   \gamma_a \gamma_b}) = \sigma (2 E_{\nu_{\mu a}} m_{\nu_{\mu}}) n_{\nu_{\mu b}},
 \end{equation}
 Using this, the mean free path (MFP) for secondary gamma rays can be calculated from 
\begin{equation}
\label{eq:intrate2}
{\lambda_{\nu_\mu\rightarrow \gamma}} = \frac{1}{\sigma (2 E_{\nu_{\mu a}} m_{\nu_{\mu}}) n_{\nu_{\mu b}}},
\end{equation}
For the scattering process involving ${\nu_{\mu a} \nu_{\mu b} \rightarrow   \gamma_a \gamma_b}$, the energy of gamma rays can be approximately equal to the energy of emitted astrophysical neutrinos. Thus, we can keep $E_{\nu_{\mu a} } \approx E_{\gamma_a}$. Using this, we have calculated the mean free path of neutrino elastic scattering as well as neutrinos annihilating into gamma rays. The results are shown in fig.~\ref{fig:mfp}. 
We can see from fig.~\ref{fig:mfp} that the black dashed line gives mean free path for neutrino for elastic scattering of incident astrophysical neutrino with CMB neutrinos for $g_{\nu_\mu}=0.01$ while the red solid line gives mean free path of neutrino for $g_{\nu_\mu}= g_{\mu}=0.01$ when the interaction of incident astrophysical neutrino with CMB neutrinos produces gamma rays. The dip in the curve shows the resonance enhancement of the cross-section at $s \approx 4 m^2_\phi$. In both cases, the mean free path for both processes lies in the range $d \gtrsim 645$ Mpc = $2 \times 10^{27}$ km. Thus, the optical depth for neutrinos traveling to earth given by $\tau = e^{-d/\lambda_{\mu\rightarrow \gamma}}$ will be around ${\cal O}(1)$. Therefore, all the gamma rays produced from neutrino scattering will reach the Earth.

\subsection{Astrophysical gamma-ray flux}
\begin{figure}[htbp]
\centering
\hskip -0.2in \includegraphics[width=0.5\textwidth, height =0.4\textwidth]{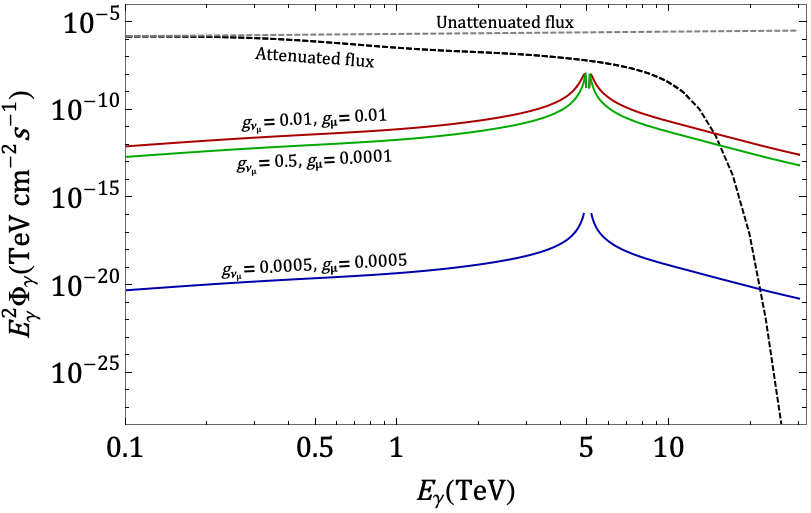}
     \caption{The gray dashed line shows the unattenuated gamma-ray flux directly coming from GRB. The black dashed line shows the attenuated gamma-ray flux directly coming from GRB. The red, green, and black solid lines show the secondary flux of astrophysical high energy gamma rays obtained from the scattering of astrophysical neutrinos with CMB neutrinos for different couplings of $g_{\nu_\mu}$ and $g_{\mu}$ respectively. The mass of the mediator has been fixed to be $m_{\phi} \approx 1$ MeV.}
     \label{fig:gammarayflux}
\end{figure}

Now we calculate the secondary flux of astrophysical photons emitted from scattering of incident astrophysical neutrinos with CMB neutrinos by using eq.~(\ref{eq:fluxfin})-(\ref{eq:intrate2}) for different values of couplings ($g_{\nu_\mu}$, $g_{\mu}$) and fixing the mass of mediator around $m_{\phi} \sim$ MeV. We assume that the background CMB photons will not acquire much energy, therefore we can keep $E_{\gamma a} \approx E_{\nu a}$. Further, we have compared our results with the unattenuated and attenuated gamma-ray flux directly coming from GRB events. As discussed above, the final expression for calculating the secondary gamma-ray flux is given by:
 \begin{equation}
 \label{eq:fluxfincopy}
 \phi^{\gamma}_{\nu_\mu} (E_\gamma) =   \ \frac{0.03 \phi^0_\gamma(E_\gamma)}{\left({\lambda_{\mu \rightarrow \gamma}}/{\lambda_{\gamma}}\right)-1} \left[e^{-d/\lambda_{\mu \rightarrow \gamma}} -e^{-d/\lambda_{\gamma}} \right].
\end{equation}
 For calculating the flux, we need to determine both $\lambda_{\mu \rightarrow \gamma}$ and $\lambda_{\gamma}$. The value of $\lambda_{\mu \rightarrow \gamma}$ has been already calculated using eq.~(\ref{eq:intrate2}). We have obtained the value of $\lambda_{\gamma}$ at different energies by using publicly available data for the optical depth of photons (calculated at redshift z =0.15) given in \cite{2011MNRAS.410.2556D} for energy up to 30 TeV. The behavior of secondary flux as a function of energy is shown in fig.~\ref{fig:gammarayflux}. We can clearly see from the figure that the flux of astrophysical gamma rays produced from neutrino sources does not attenuate at high energies even though the magnitude of flux is lower than the direct gamma-ray flux at lower energies. The results are shown for three different benchmark values of the coupling of $g_{\nu_\mu}$ and  $g_{\mu}$ respectively. The choice of benchmark couplings has been motivated by Hubble tension requirement as well as the the allowed region of $g_{\mu}$ from experimental results of $(g-2)_{\mu}$  etc. We will discuss the specific choice of coupling parameters in the next subsection.

\begin{figure*}[htbp]
  \centering
  \hspace*{-1cm} 
 \subfigure[ $~g_{\nu_\mu} = g_{\mu}$]{\includegraphics[width = .5\textwidth,keepaspectratio]{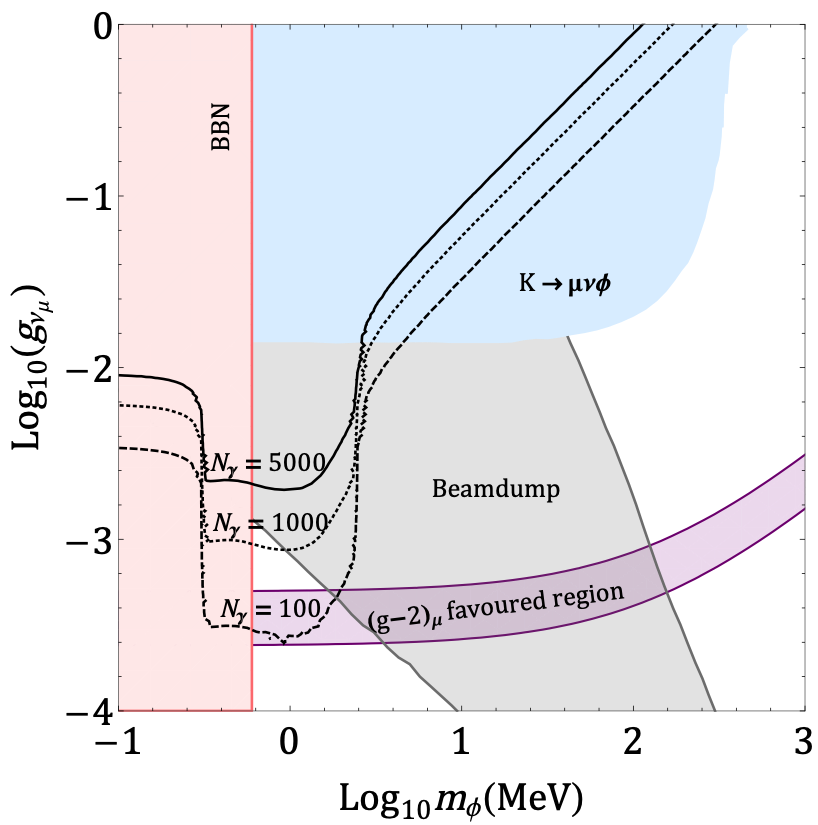}} \quad
 ~~~~\subfigure[$~g_{\nu_\nu} = g_{\mu} $ (including Hubble tension constraints)]{\includegraphics[width = .5\textwidth,keepaspectratio]{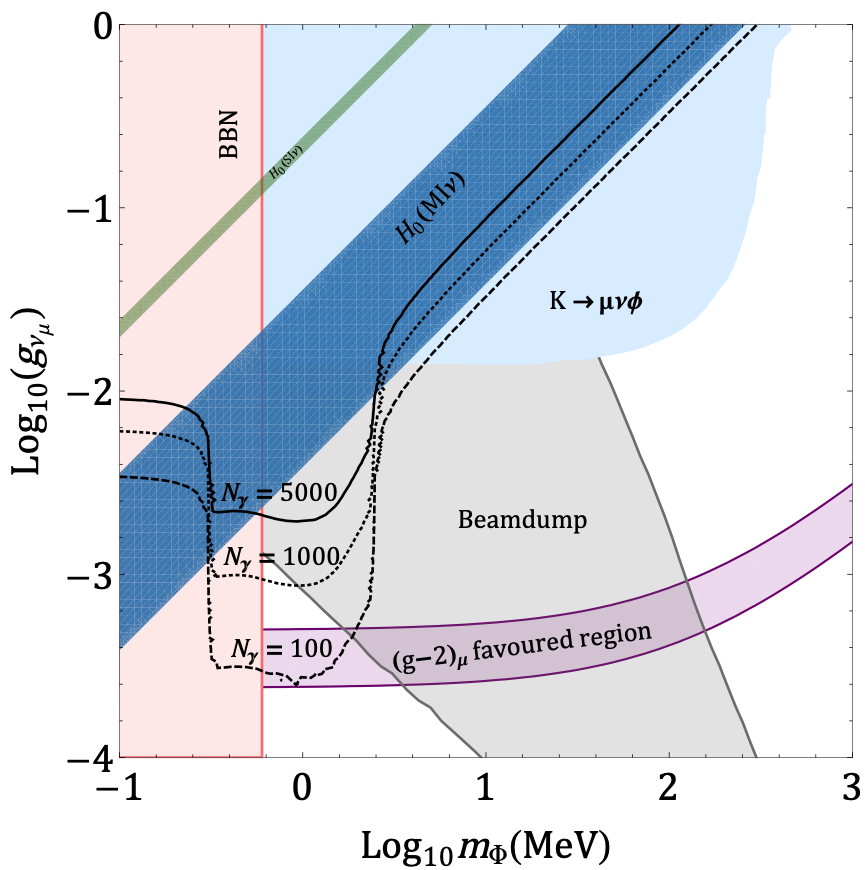}}  
  \caption{
The dashed, dotted, and solid black curves correspond to parameter space required to observe $N_{\gamma} =100$, $N_{\gamma}= 1000$, $N_{\gamma} = 5000$ events respectively. The light-red shaded region represents the parameter space ruled out by constraints from BBN~\cite{Blinov:2019gcj}. The light-blue shaded region shows the excluded parameter space from the constraint on the branching ratio of kaon decay: $K \rightarrow \mu \nu_{\mu} \phi$~\cite{Krnjaic:2019rsv}. The gray shaded region excludes the parameter space from beam-dump experiments~\cite{Depta:2020wmr}.  In the right-hand side figure, the blue and green shaded bands correspond to MI$\nu$ and SI$\nu$ region allowed by Hubble tension constraints~\cite{Lancaster:2017ksf}.  In the left-hand figure, we can see that the small amount of parameter space available for $N_\gamma =100$ events is also consistent with the allowed region from $(g-2)_{\mu}$~\cite{Liu:2021kug}. 
  }
   \label{ps1}
\end{figure*}

\begin{figure*}[htbp]
  \centering
  \hspace*{-1cm} 
 \subfigure[ $~g_{\nu_\nu} \neq g_{\mu}, g_{\mu} = 5 \times 10^{-4}$]{\includegraphics[width = .5\textwidth,keepaspectratio]{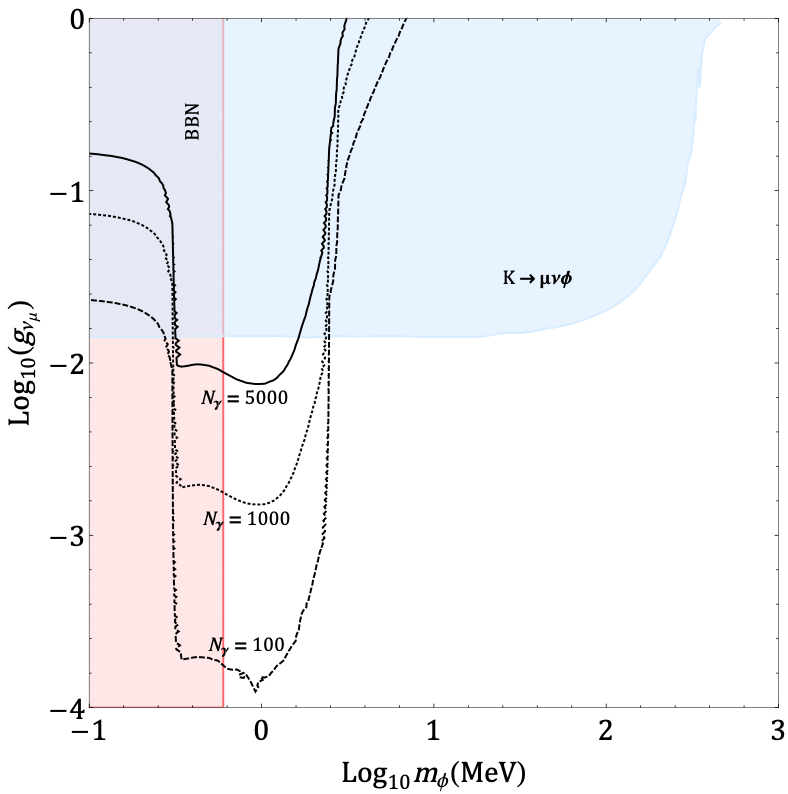}} \quad
  \subfigure[$~g_{\nu_\nu} \neq g_{\mu}, g_{\mu} = 5 \times 10^{-4}$(including Hubble tension constraints)]{\includegraphics[width = .5\textwidth,keepaspectratio]{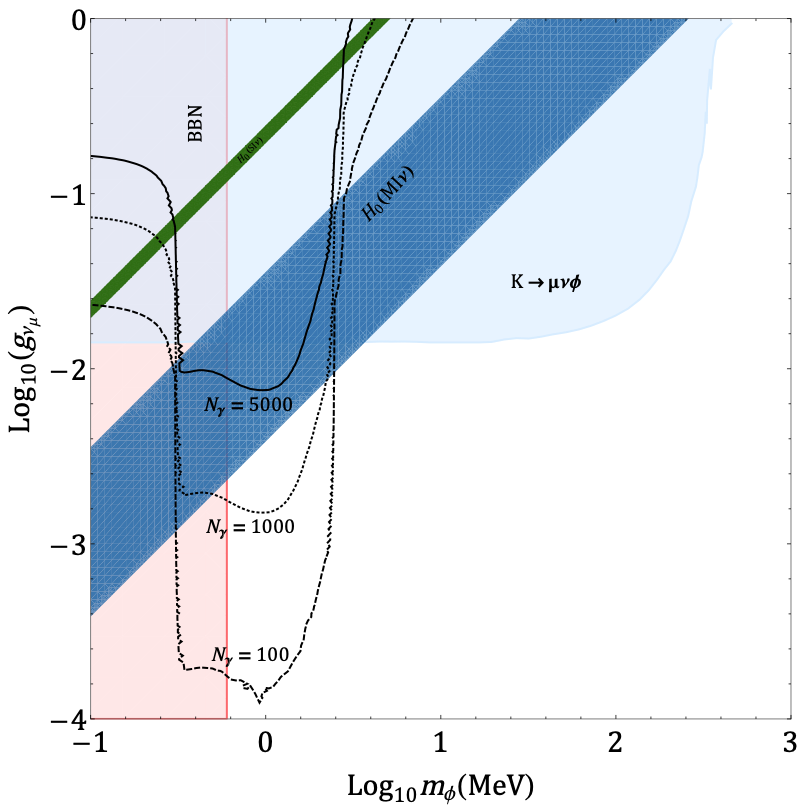}}  
  \caption{
The dashed, dotted, and solid black curves correspond to parameter space required to observe $N_{\gamma} =100$, $N_{\gamma}= 1000$, and $N_{\gamma} = 5000$ events respectively. The light-red shaded region represents the parameter space ruled out by constraints from BBN~\cite{Blinov:2019gcj}. The light-blue shaded region shows the excluded parameter space from the constraint on the branching ratio of kaon decay: $K \rightarrow \mu \nu_{\mu} \phi$~\cite{Krnjaic:2019rsv}. In the right-hand side figure, the blue and green shaded bands correspond to MI$\nu$ and SI$\nu$ region allowed by Hubble tension constraints~\cite{Lancaster:2017ksf}.  
  }
   \label{ps2}
\end{figure*}

\subsection{Parameter Space of neutrino self-interaction coupling vs mass of the mediator}
In this subsection, we estimate the total number of secondary gamma-ray events observed from the scattering of astrophysical neutrinos with CMB neutrinos. The number of events can be computed by multiplying the flux with an effective cross-section area and observation time. Using an effective area of 1 km$^2$ and observation time window of 2000 sec~\cite{Smirnov:2022suv}, the number of events in the energy range $E_{\gamma} \sim (1-30)$ TeV  can be calculated from
\begin{equation}
\label{eq:numberofevents}
N_{\gamma} = \int^{30{\rm TeV}}_{1 \rm TeV} \phi^{\gamma}_{\nu_\mu}(E_\gamma) \ dE_{\gamma} \ dA    \ dt, 
\end{equation}
where $\phi^{\gamma}_{\nu_\mu} (E_\gamma)$ corresponds to flux given in eq.~(\ref{eq:fluxfincopy}).
As $N_\gamma$ depends on the self-interaction neutrino coupling ($g_{\nu_\mu}$) and mass of the mediator ($m_{\phi}$), we can use eq.~(\ref{eq:numberofevents}) in order to constrain the parameter space of $g_{\nu_\mu}$ as a function of $m_{\phi}$. Thus, we have obtained $g_{\nu_\mu}-m_{\phi}$ parameter space by considering the different observed number of events. In particular, we consider three cases for the observed number of events: (i) $N_{\gamma} = 100$ (ii) $N_\gamma = 1000$ (iii) $N_\gamma = 5000$. The values of $g_{\nu_\mu}$ and $g_{\mu}$ depend on the underlying model of BSM physics. As we do not consider any UV complete model of BSM physics, we are assuming that the values of both couplings can either be different or the same. Therefore, while obtaining the parameter space, we have considered two possibilities: (a) $g_{\nu_\mu} =  g_{\mu}$, (b) $g_{\nu_\mu} \neq g_{\mu}$. For the second case, we have fixed the value of $g_{\mu}$ allowed by constraints from the recently measured value of $(g-2)_\mu$~\cite{Liu:2021kug}. The results are shown in figs.~\ref{ps1} and \ref{ps2} for both cases along with parameter space allowed by Hubble tension requirement and ruled out by other cosmological/ laboratory constraints. The dashed, dotted, and solid black curves in both figures 
show the parameter space required to observe $N_{\gamma} =100$, $N_{\gamma}= 1000$, $N_{\gamma} = 5000$ events respectively.  Below, we discuss all other constraints shown in figs.~\ref{ps1} and \ref{ps2}.
\paragraph{Constraints from Hubble Tension:} 
 As discussed in~\cite{Lancaster:2017ksf}, the strength of self-interacting neutrino required to get the right value of Hubble constant can be categorized in two regimes, dubbed as strong-interacting neutrino (SI$\nu$) and moderately interacting neutrino (MI$\nu$). The values of $G_{\mathit{eff}}$ in both regimes are given as~: 
\begin{equation}
G_{\mathit{eff}} = \begin{cases}
            (4.7^{+0.4}_{-0.6}{\rm~MeV})^{-2},~{\rm~SI\nu}  \\
             (89^{+171}_{-61}{\rm~MeV})^{-2},~{\rm~MI\nu}. 
              \end{cases}
\end{equation} 
The resulting parameter space along with Hubble tension constraints is shown separately in figs.~\ref{ps1}\textcolor{red}{(b)} and \ref{ps2}\textcolor{red}{(b)} for both cases by superimposing allowed range of $G_{\rm eff}$ as blue-shaded and green-shaded band respectively.

\paragraph{Various other Cosmological and laboratory constraints:} The new interaction between neutrinos and scalar mediator allows the light scalar mediator to be in thermal equilibrium before the onset of neutrino decoupling and affect the total number of relativistic degree of freedom ($\Delta N_{\rm eff}$) present in the universe. Therefore, the requirement $\Delta N_{\rm eff} \lesssim 0.5$ puts a bound on the mass of real scalar mediator to be $m_{\phi} \gtrsim 0.16$ MeV~\cite{Blinov:2019gcj}. The ruled-out region is shown as a light-red shaded band in both figs.~\ref{ps1} and \ref{ps2}. The constraints from the laboratory originate from the possible decay channel of kaon to the light scalar given by $K \rightarrow \mu \nu_\mu \phi$. Therefore, the experimental bounds on the kaon decay rate also put a bound on the coupling $g_{\nu_\mu}$ for $m_{\phi} \leq m_{\mu}$~\cite{Krnjaic:2019rsv} in fig.~\ref{ps1} and \ref{ps2}. The ruled-out parameter space from this bound is shown as a light-blue shaded region in both figures.
In the presence of a new scalar mediator coupled to the muon, the experimentally allowed value of $\Delta a_{\mu} =(g-2)_\mu/2$ puts a bound on the coupling parameter $g_{\mu}$. Therefore, for the case of $g_{\nu_\mu} =  g_{\mu}$ as shown in figs.~\ref{ps1}\textcolor{red}{(a)} and \ref{ps1}\textcolor{red}{(b)}, we have considered constraints from the updated value of $(g-2)_{\mu}$~\cite{Liu:2021kug}. The allowed region is shown as purple-shaded band in figs.~\ref{ps1}\textcolor{red}{(a)} and \ref{ps1}\textcolor{red}{(b)}. Similarly, the interaction coupling $g_\mu$ also gets constrained from beam dump experiments~\cite{Depta:2020wmr}. The gray shaded region in figs.~~\ref{ps1}\textcolor{red}{(a)} and \ref{ps1}\textcolor{red}{(b)} excludes the parameter space from beam-dump experiments~\cite{Depta:2020wmr}. For $g_{\nu_\mu}\neq g_{\mu}$, we have already fixed the value of $g_{\mu} = 5 \times 10^{-4}$ which is consistent with the allowed experimental value of $\Delta a_{\mu}$. Thus, the bound from $(g-2)_{\mu}$ does not exist on $g_{\nu_\mu}$ in figs.~\ref{ps2}\textcolor{red}{(a)} and \ref{ps2}\textcolor{red}{(b)}. The bound from beam dump experiment will also not apply in the second case shown in figs~\ref{ps2}\textcolor{red}{(a)} and \ref{ps2}\textcolor{red}{(b)}.

Finally, we realize that a tiny amount of parameter space remains available for $N_{\gamma} = 100$ events for the case of $g_{\nu_\mu}=g_{\mu}$, which is also consistent with the allowed region of  $(g-2)_{\mu}$. In this case, the region favored by Hubble tension constraints is ruled out by all other constraints. For $g_{\nu_\mu} \neq  g_{\mu}$, the parameter space remains available for $N_{\gamma} \sim (100-5000)$ events. Interestingly, some of the parameter space for $N_{\gamma}= 5000$ events is also consistent with the MI$\nu$ range allowed by Hubble tension constraints and free from other laboratory and cosmology constraints. Thus, we conclude that the scattering of an astrophysical neutrino with CMB neutrinos can be the origin of high-energy astrophysical events observed from GRB, and some of the allowed region of $g_{\nu_\mu}$ is also consistent with allowed values from Hubble tension constraints and $(g-2)_\mu$ discrepancy.

\section{Concluding Remarks}
In the last few years, neutrino astronomy has turned out to be extremely useful in providing a  more comprehensive understanding of the universe. In this work, we have emphasized the role of neutrinos in explaining high-energy gamma-ray events observed in GRB221009A through non-standard self-interaction of neutrinos. The model of self-interacting neutrinos is primarily motivated in cosmology for resolving the discrepancy between the value of the Hubble constant measured by CMB observations and low redshift experiments. The inclusion of the same can delay the epoch of neutrino decoupling and also modify the CMB power spectrum obtained in the $\Lambda$CDM model. As a result of this, the comparison of the modified CMB power spectrum with the measured CMB power spectrum allows a high value of Hubble constant in the $\Lambda$CDM model, thus reducing the Hubble tension. The detailed CMB analysis of the same allows a very specific range of self-interaction coupling vs. mass of the mediator. In this work, we assume that the same interaction of the scalar boson with neutrinos can also produce secondary gamma rays from the scattering of astrophysical neutrinos with CMB neutrinos if the new scalar mediator interacts both with muon neutrinos and its leptonic partner. The interaction of scalar with muons is already motivated by the discrepancy between the theoretical and experimental values of $(g-2)_{\mu}$. Basically, the interaction of an astrophysical neutrino with CMB would produce a scalar boson, which can further decay into a high-energy astrophysical photon and a CMB photon at a one-loop level through the interaction of scalar mediator with muons.

By considering a toy model of a light scalar interacting with muon neutrinos and muons, we have calculated the flux of high-energy astrophysical gamma rays produced through such a process and show that we can obtain the required number of gamma-ray events produced by GRB221009A without having the same to be attenuated while traveling to earth. In fact, some of the resulting parameter space of self-interacting neutrino coupling is also in agreement with the parameter space obtained from Hubble tension requirements, allowed $(g-2)_\mu$ region and free from other laboratory and cosmology constraints as well. Thus, the strong self-interaction between neutrinos in the astrophysical environment can explain the origin of high-energy gamma-ray events observed in GRB221009A. In the future, as CMB observations become precise, the constraints on the allowed region of self-interaction between neutrinos will become more stringent. Therefore, it would be interesting to study the consequences of the same in astrophysical processes. In fact, the exotic high-energy astrophysical gamma rays emitted from the scattering of neutrinos can act as a complementary probe to unveil the source of high-energy astrophysical neutrino events.

\section*{Acknowledgments}

 MD  would like to acknowledge support through the  DST-Inspire Faculty Fellowship of the Department of Science and Technology (DST), Government of India under the Grant Agreement number: IFA18-PH215.  MD  would also like to thank the organizers of ``Annual Theory Discussion days-2023" held at PRL Ahmedabad, India where preliminary results of this work were presented.
\bibliography{bibliography_1}
\bibliographystyle{unsrt}

\end{document}